\documentclass{pasj00}

\RelativeFigurePath{}

\begin{document}
\SetRunningHead{Costa, Sim\~oes, Pinto and Melnikov}{Burst Analysis
  with 3D Loop Models}

\title{Solar Burst Analysis with 3D Loop Models}
\author{
   Joaquim Eduardo Rezende \textsc{Costa}\altaffilmark{1}
   Paulo Jos\'e de Aguiar \textsc{Sim\~oes}\altaffilmark{2}
   Tereza Satiko Nishida \textsc{Pinto}\altaffilmark{1,3}
   and
   Victor Fedorovich \textsc{Melnikov}\altaffilmark{4}
}

\altaffiltext{1}{Astrophysics Division, National Institute for Space
  Research, Av. Astronautas, 1758, 12227-010 S\~ao Jos\'e dos Campos,
  Brazil} \altaffiltext{2}{SUPA, School of Physics and Astronomy,
  University of Glasgow, Scotland} \altaffiltext{3}{Nobeyama Solar Radio
  Observatory, National Astronomical Observatory of Japan, Minamimaki,
  Minamisaku, Nagano 384-1305} \altaffiltext{4}{Central Astronomical
  Observatory at Pulkovo of the Russian Academy of Sciences,
  Saint-Petersburg 196140, Russia}

\email{jercosta@das.inpe.br}
\KeyWords{Sun: radio radiation, Sun: flares, Sun: magnetic fields,
  Sun: atmosphere, catalogs} %

\maketitle

\begin{abstract}
  A sample of Nobeyama flares was selected and analyzed
  using loop model for important flare parameters. The model for the
  flaring region consists of a three dimensional dipolar magnetic
  field, and spatial distributions of non­thermal electrons. We
  constructed a database by calculating the flare microwave emission
  for a wide range of these parameters. Out of this database with more
  than 5,000 cases we extracted general flare properties by comparing
  the observed and calculated microwave spectra. The analysis of NoRP
  data was mostly based in the center ­to ­limb variation of the flare
  properties with looptop and footpoint electron distributions and for
  NoRH maps on the resultant distribution of emission. One 
important aspect of this work is the comparison of the analysis of a flare 
using an inhomogeneous source model and a simplistic homogeneous 
source model.  Our results show clearly that the homogeneous source hypothesis is not 
appropriate to describe the possible flare geometry and its use can easily 
produce misleading results in terms of non-thermal electron density and magnetic 
field strength. A center darkening of flares was also obtained as a geometrical property of 
the loop-like sources
\end{abstract}

\section{Introduction}
Solar flares are associated with magnetic loops or arcades, where
accelerated particles move along the field, and can be trapped due to
magnetic mirroring or precipitate into the chromosphere, where they
produce hard X-rays by bremsstrahlung. The radio/microwave flare
emission is produced by electrons moving in the magnetic field due to
the gyrosynchrotron mechanism. It has been widely demonstrated that
the characteristics of this radiation (spectrum, polarization, spatial
distribution) are strongly dependent on the magnetic field strength
and geometry, as well as the properties of the non-thermal electrons,
such as energy, pitch-angle and spatial distributions
\citep{Alissandrakis:1984,KleinTrottet:1984, Melnikov etal:2002, 
FleishmanMelnikov:2003, 2006A&A...453..729S, Reznikova_etal:2009, 2010SoPh..266..109S}
. High spatial resolution flare observations in microwaves with the
Nobeyama Radio Heliograph (NoRH), Very Large Array (VLA) and Owen
Valley Solar Array (OVSA) and in X-rays with Yohkoh and Reuven Ramaty
High Energy Solar Spectroscopic Imager (RHESSI) have shown a great
variety of source morphologies, interpreted as magnetic arcades,
interaction between two or more loops, and many others. Nevertheless,
it seems that many flares can be associated with a single magnetic
loop, i.e. conjugated footpoints connected by a coronal loop
 (e.g. \cite{Melnikov:2006, Tomczak:2007,Tzatzakis:2008}).

The lack of knowledge on the coronal magnetic field usually leads to
simple homogenous source model to analyze the emission, although
simple field geometries were proposed to model the observed microwave
emission of specific flares \citep{Nindos:2000,Kundu:2001}.  To
describe a spatially varying flare source, it is necessary to define
the strength and geometry of the magnetic field, spatial distribution
of the non-thermal electrons, and to a lesser importance for microwave
emission, the thermal plasma density and temperature, which can be
relevant to the free-free absorption and Razin effect.  To improve our actual 
knowledge from the flare homogeneous modeling 
we constructed a database of models using a
three dimensional dipolar loop geometry and searched for the ranges of
parameters to describe the observables. The implementation of a
dipolar loop geometry is a great improvement from the homogeneous
flare model in terms of the description of the flare source.

The motivation of our work was the construction of a models database
to be a framework for a joint analyzes of microwave solar flares. We
searched for the set of geometrical conditions to a magnetic bottle
that reproduce the statistical properties of the spectra observed by
the Nobeyama Radio Polarimeter (NoRP)  \citep{nakajima85,torii79} and the brightness distribution
maps observed by the Nobeyama Radioheliograph (NoRH) \citep{nakajima94, takano97}. The aim of the work
was not to construct a new interactive tool for flare analysis, but to
present an ensemble of 
models with emission characteristics similar of those
observed flares to infer physical 

parameters of flare loops from this set and evaluate
the risk of getting misleading results.

Our approach can be considered as a first order approximation of the realistic
distribution of parameters. Considering actual understanding of flare
emission in microwaves about emission mechanisms, geometry, particle
distribution in energy and space and the flare ambient (density and
magnetic field), we constructed the template of our model, in a
similar fashion to other interesting works of
\cite{2010SoPh..266..109S}, \cite{Nita:2009uw,2011ApJ...742...87K} and
references therein. To analyze the flare properties of our database
we calculated more than 5,000 models with a range of parameters capable to
reproduce the observed flare characteristics, such as flux densities,
spectrum characteristics and source sizes.


\section{Heliographic distribution of flares}
\begin{figure*}
  \begin{center}
    \FigureFile(70mm,80mm){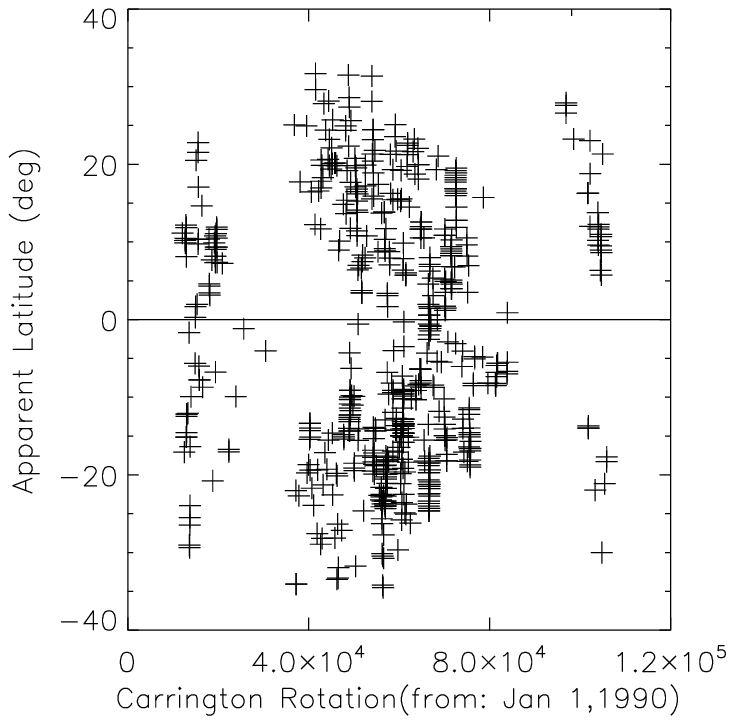}
    \FigureFile(70mm,80mm){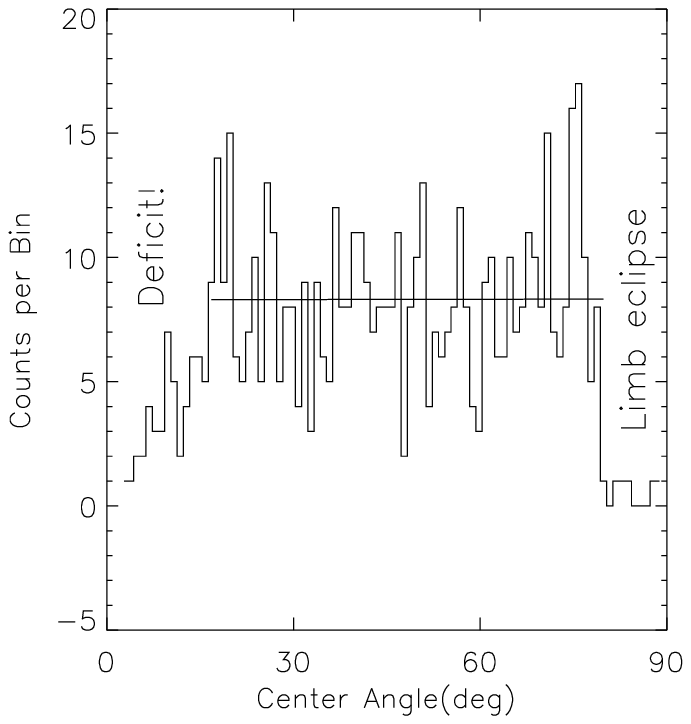}
  \end{center}
  \caption{Left frame: Butterfly diagram with apparent heliographic latitude
    (without the correction for the tilt of the solar equator to the
    ecliptic). Right frame: Histogram of center angle for NoRH flares.}
\label{fig:butter}
\end{figure*}

Figure \ref{fig:butter}a shows NoRH flare's heliographic coordinates for
600 observed events in all possible latitudes as known from the
butterfly diagram of the active regions. It is noticeable the lack of
flares observed at disk center \citep{2010Costa}.

The distribution of flares in apparent latitude (i.~e., in relation to the visual center) also shows a strong decrease of the flares at the
disk center, as shown in the Figure \ref{fig:butter}b. In this
distribution we used the central angle as the position coordinate of
the flare. The central angle is defined as the angle between the
normal of flare position and the normal to the disk center. This
defines 0\degree and 90\degree as the disk centre and solar limb,
respectively. Although we see flares near the visual equator the
center to limb distribution of flares is not homogeneous. Around
80\degree there is an abrupt decrease in the number of flares, most
likely due to imprecision in measuring the position close to the limb
or to auto eclipse. Center angles smaller than 20\degree show a
smaller number of flares, i.~e., at disk center there is a deficit of
events. This result is a strong motivation for this work, as position
and geometry of flaring loops may explain these results. There is a
strong dependence of the gyrosynchrotron emission with the angle
between the magnetic field and the line of sight ($\theta$). If most
of the emission comes from the footpoints and legs of the magnetic
bottle, at the disk center the angle of the magnetic field, $\theta$,
tends to zero and the emission becomes negligible, resulting in the
observed effect of missing flares.

\section{Model Description}

The canonical geometry for a solar flare is a magnetic bottle where
the accelerated particles are trapped by magnetic mirroring and emit
microwave radiation.  The geometry in the numerical model was based on
an analytic dipole magnetic loop, which can be placed in any position
observed by NoRH. The magnetic field lines to be populated by thermal
plasma and nonthermal electrons are free parameters restricted only by
observed source sizes in 17 GHz at NoRH brightness maps. Finally,
microwave images are computed in any chosen frequency.  Ambient
density and temperature were filled using plan-parallel atmospheric
model, with distance from the center of the Sun. The non-thermal
electron density was filled with distance from the top of the
structure and considering the total length of the line. 
\subsection{Magnetic field model}
\begin{figure}
 \begin{center}
    \FigureFile(90mm,80mm){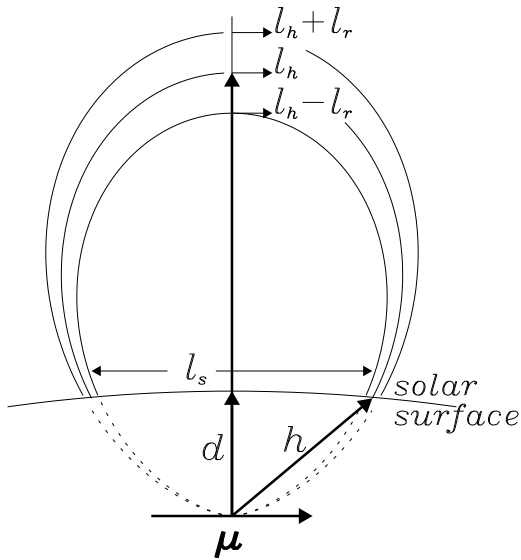}
 \end{center}
  \caption{Magnetic model.}
\label{fig:dipole}
\end{figure}

To define the flaring loop magnetic field we used a dipole magnetic
field described by:
\begin{equation} \label{eq:B} B=\frac{ 3( {\boldsymbol \mu \boldsymbol
      r}){\boldsymbol r}-r^2 {\boldsymbol \mu}}{r^5}
  \label{eq:dipole}
\end{equation}
where the dipole is in the origin of coordinates, the
$\boldsymbol r $ is the position vector of any voxel in space and
$\boldsymbol \mu$ is the dipole magnetic moment with absolute value
calculated from the model condition of a given maximum magnetic field
at the loop footpoint (see figure \ref{fig:dipole}). The figure
\ref{fig:dipole} shows a magnetic dipole with a moment $\mu$ paralell
to the tangent of the solar surface. The dipole is placed below the
solar surface at depth $d$. The maximum magnetic field strength is a
free parameter for the lowest level of the loop volume (B($h$)), where
$h$ is the modulus of the position vector $\boldsymbol h$ of the
footpoint. The flaring volume is constructed around the central field
line with a circular cross-section with radius at the apex of $l_{r}$
as shown in figure \ref{fig:dipole}. Thus, the geometrical free
parameters are the loop height ($H=l_{h}-d$), feet separation
($l_{s}$) and apex radius ($l_{r}$). The dipole depth ($d$) below the
surface is calculated from the selected $H$ and $l_{s}$ which in turn
changes the mirror ratio (B($l_{h})$/B($h$)).
The loop can be placed in any position on the Sun, by defining its
latitude and longitude, and with any azimuth angle i.~e. the rotation
in relation with the solar equator. This geometry is defined inside a
$32\times 32\times 32$ voxels (volume units) cube. The $Z$ direction
points to the observer whatever is the solar position. The viewing
angle $\theta$, i.~e. the angle between the magnetic field direction
at each point and the observer line-of-sight, is calculated from the
dipole field lines.

\subsection{Spatial distribution of particles}
\subsubsection{Non-thermal electrons}
The description of the distribution functions of the non-thermal
electrons in solar flares is still one of the critical
unanswered. Several processes such as Coulomb collisions,
wave-particle interactions, return currents, magnetic trapping can
influence the transport of electrons, and this still is subject of
intense investigation. However, a detailed description of such
transport processes to evaluate the spatial distribution of electrons
is out of the scope of our analysis; moreover, the computing time of
these solutions is known to be in a non practical scale for a big
number of calculations. To restrict the computing time, we defined
empirical functions for the electron distributions in energy and
space. 

\begin{figure}
 \begin{center}
    \FigureFile(90mm,80mm){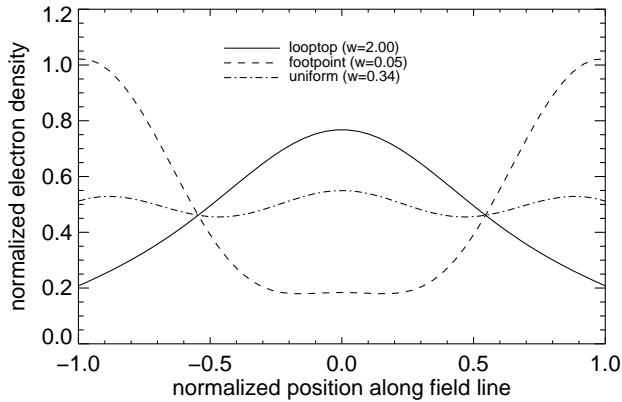}
 \end{center}
 \caption{Weighting function for the three distributions used in the models defined by the equation \ref{eq:par}}. 
\label{fig:distr}
\end{figure}

For the energy distribution, we adopted a single power law
distribution (with power index $\delta$ as a free parameter), in the
energy range from 10 keV to 100 MeV, in a isotropic pitch-angle
distribution. To define the spatial distribution of the electrons
$N(s)$ we used an empirical weighting function parametrized by a
constant $w$ as follows:
\begin{equation} \label{eq:par}
     N(s)=N_{\mathrm{el}}\frac{(2 s^4 + w) e^{-2 s^{2}}}{a+b w}
\end{equation}
where $s$ is the normalized (unit size for all field lines)
distance along the field line from the loop apex and $w$ is the
selection parameter for three distributions: looptop, footpoints or
homogeneous concentrations, given by $w=2.0$, 0.05, 0.34,
respectively (see figure \ref{fig:distr}). The normalized $s$ resulted in uniform distribution
  along the cross section of the loop. The spatial distribution is
symmetrical in relation to the apex. Once the geometrical
  parameters of the loop were defined, each voxel inside the
  volume has a field line determined by equation \ref{eq:dipole} that
  passes through its center, thus the height $l_i$ of that line can be
  calculated. If $l_h-l_r < l_i < l_h+l_r$ (Figure \ref{fig:dipole}) the
  voxel associated with the line is marked as internal to the loop
  volume. This procedure is repeated for all the voxels inside the
  volume, determining the discrete loop geometry, with a defined
  number of voxels, and thus the total volume of the loop. Then, the
  normalization is numerically computed after applying the equation
  \ref{eq:par} in each normalized voxel distance $s$. Parameters $a$
and $b$ normalize the weighting function to maintain the total number
of trapped electrons inside the loop volume $V_{\mathrm{loop}}$ equal
to the user defined non-thermal electron number,
i.~e. $N_{tot}=N_{\mathrm{el}}$ $\times V_{\mathrm{loop}}$.

\begin{figure}
 \begin{center}
    \FigureFile(90mm,80mm){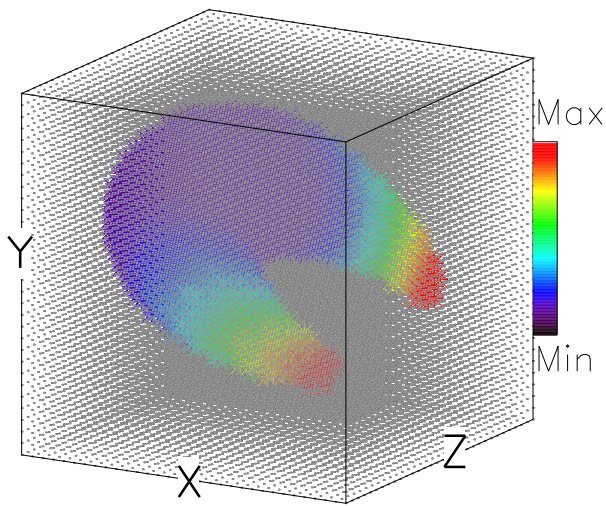}
 \end{center}
 \caption{Apparent loop geometry at N30W85 solar heliographic
   coordinates. The color coded image shows the electron number density $N_{el}$, with a 
   footpoint concentration of electrons. 
 Coordinates $[X,Y,Z]$ point to solar East, North, Observer direction,
   respectively.}
\label{fig:loop}
\end{figure}

Figure \ref{fig:loop} shows
one case of magnetic loop centered on the heliographic coordinate
N30E85. The axis connecting the two footpoints was rotated by
15\degree in relation to the solar equator. Figure \ref{fig:loop} shows the
normalized field line length coordinate ($s$) (zero is the loop apex
and one is the footpoint). The color code indicates the distance of the regions to 
be filled with the nonthermal electron distributions, which simulate looptop or
footpoint concentrations.
Figure \ref{fig:loop} shows an example of electron number density ($N_{el}$) distribution, 
where we used the normalized field line length coordinate ($s$) (zero is the 
loop apex and one in the footpoint) to fill the loop according to equation \ref{eq:par} 
for a footpoint concentration of electrons (w=0.05). \\

\subsubsection{Ambient plasma}

For the ambient thermal plasma, the temperature was fixed at a high 
value to minimize the influence of free-free emission typical
value during flares at $10^{8} K$, that minimizes the free-free contribution;
specifically absorption. The density has an exponential decay from
10$^{10}$ cm$^{-3}$ to 10$^{8}$ cm$^{-3}$. With these figures the 
contributions of bremsstrahlung and the Razin effect in the range of 
frequency considered in this work were negligible. 

\subsection{Microwave emission mechanisms}
The gyrosynchrotron emission $j_\nu$ and self-absorption $k_\nu$
coefficients were numerically calculated for the two propagation modes
(ordinary and extra-ordinary) using the full formalism
\citep{1969ApJ...158..753R}. The numerical code was optimized for
parallel processing in multi-core processors, and approximations for
higher order of the Bessel functions \citep{1971AuJPh..24...43W} were
implemented to improve the computing speed without any significant
loss in numerical precision. The effect of 
Razin suppression is
included in our calculations since it is a natural consequence when
considering the full gyrosynchrotron formalism, where the production
of the radiation is strongly suppressed for frequencies below the
Razin frequency 
$\nu_R={2 \nu_p^2}/(3 \nu_B sin~\theta)$ which  relates the gyrofrequency $\nu_B$ and
the plasma frequency $\nu_p$. 

The free-free emission and absorption from the thermal plasma are also
included in the code, although their effects are not significant for
the range of plasma parameters considered here. Also, the Razin effect
is not effective 
in the range of parameters and frequencies here considered. In other
words, the construction of this database aimed to reproduce the
characteristics of the gyrosynchrotron emission with minimum Razin
suppression and free-free contribution.

We noted that for $90\%$ of the Nobeyama sample of flares, the spectral indices in the range of 
frequencies below the 
peak frequency (between 3.75 and 9.4 GHz) are  smaller than 2.5 (the expected index for 
gyrosynchrotron emission of non-thermal electrons in 
homogeneous source in the optically thick part of the spectrum  \citep{1985ARA&A..23..169D}). 
We emphasize that our model bank is only a first order approximation for the Nobeyama bursts, 
because of the high number of unknowns that was shortened here for the sake of the decrease the processing time.

The frequencies of the model calculations are 3.7, 9.4, 17, and 34
GHz, selected considering NoRP and NoRH observations. These
frequencies allow to study gyrosynchrotron spectrum without
contamination from low frequency plasma emission mechanisms
\citep{2004ApJ...605..528N}. We did not include 80 GHz as it is not
always available in the observational data. The exclusion of such high
frequency also speeded up the computing time.

\subsection{Solution of the radiative transfer in a 3D source}
The ray path in our model is treated as straight lines as in
vacuum-like ray trace. This assumption is valid since the refraction
indices for the range of frequencies considered here are close to
unity for the entire range of source parameters used in the
calculations. 

The voxels in our 3D source model were constructed small enough to be
considered a homogeneous unit of volume, inside of which all physical
variables are uniform. Thus, the analytic solution of the radiative
transfer equation can be applied to the voxels at some level
$\ell$ with an input of radiation from the level $\ell-1$ below. The
specific intensity $I_\nu$ emerging at each pixel in the $XY$ plane of
the source is obtained by:

\begin{eqnarray}
\label{transf}
  S_{\nu \ell}&=&\frac{\jmath_{\nu \ell}}{k_{\nu \ell}} (1-e^{-k_{\nu\ell} \Delta L}) \\
  I_{\nu} &=& \sum_{\ell=2}^{n} [S_{\nu n} + I_{\nu(\ell-1)} e^{-k_{\nu\ell} \Delta L}]
\end{eqnarray}

where $\ell$ is the integration level in the direction of the observer
$Z$, from one (cube bottom) to $n$ (cube top), $\jmath_{\nu \ell}$ is
the gyrosynchrotron emission coefficient and $k_{\nu \ell}$ the
absorption at frequency $\nu$ and level $\ell$. The voxel size is
$\Delta L$. This procedure is repeated for each frequency, producing
emission maps for both modes of propagation. The total flux density
maps were obtained by summing the intensity of both modes, then
multiplying by the angular size of the pixels. The spatially
integrated flux density is easily calculated by summing up all the
pixels in each map.

\section{Model database}
The database was constructed with the 15 variable parameters previously described. In this actual version the selected frequencies were [3.75, 9.4 17.0, 35.0] GHz, the number of voxels is 32$\times$32$\times$32, $l_{r}=0.007 R_{\solar}$ (where $R_{\solar}$ is the solar radius), $l_{s}=0.02 R_{\solar}$, $H=0.007 R_{\solar}$, ambient density exponentially decaying with height from $10^{10} - 10^{8} cm^{-3}$, energy range of the power law distribution of nonthermal electrons from $10 keV$ to $100 MeV$, homogeneous pitch angle distribution and constant ambient temperature at $10^{8} K$. 
The $w$ constant [0.05, 0.34, 2.] in equation \ref{eq:par} was homogeneously distributed among these three values and the loop azimuth was also equally distributed among [-60\degree,-15\degree,15\degree,60\degree]. The loop heliographic position was randomly selected from about 600 positions of flares observed by NoRH. The nonthermal electron number density $N_{el}$, the power law index of the energy distribution $\delta$ and the magnetic field at the footpoint $B_{foot}$ were varied in the ranges shown in table \ref{tab:range}. The ranges of parameters were dynamically chosen to reflect the range of flux densities observed by NoRP. The total volume was kept constant as a decision to decrease the parameters' space but also considering results of the NoRH flares at 17 GHz. The actual version of the database has 5,000 computed models and size of 10 GB. 

\begin{table}
  \caption{Range of parameters}\label{tab:range}
  \begin{center}
    \begin{tabular}{lrr}
      \hline
       Parameter & Minimum & Maximum \\  
      \hline
      B$_{\mathrm{foot}}$ (G) & 1200 & 1800 \\
      N$_{\mathrm{el}}$ (cm$^{-3}$) & $10^{6}$ & $2.5\times 10^{7}$ \\
      $\delta$ & $1.8$ & $2.4$ \\
      \hline
    \end{tabular}
  \end{center}
\end{table}

\section{Some results}
\begin{figure*}
 \begin{center}
    \FigureFile(170mm,120mm){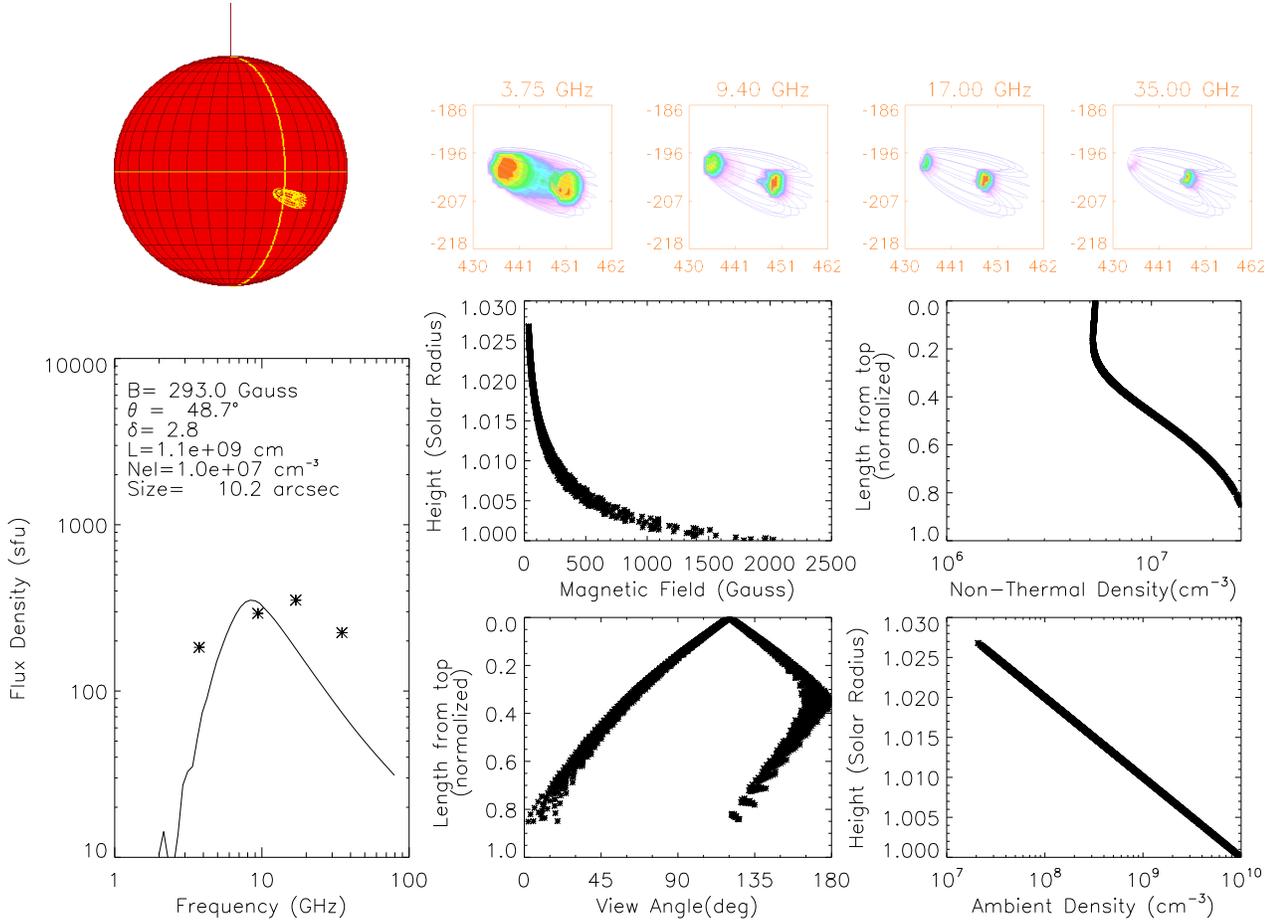}
 \end{center}
  \caption{Model result. Example of one element of the data bank.}
\label{fig:result}
\end{figure*}

A snapshot from a model can be seen in the figure
\ref{fig:result}. The Sun is shown at the upper left with the loop
over-plotted (the size ten times enlarged for better
visualization). Four images of calculated brightness are shown on the
upper part. The scale of the images are in arcsec from the Sun
center. The integrated flux density for these images are plotted as
asterisks on the spectrum at left of the figure. The over-plotted
continuum line is the spectrum for a homogeneous source for the
parameters shown on the figure. These parameters were calculated from
our data cubes to be representative (effective numbers) of our
inhomogeneous model. They were calculated using the image with the
highest flux density (in this case 17 GHz) with the brightness being
the weighting function for all spatially varying parameters.

The source size was calculated to give the same peak flux density of
the inhomogeneous model. The result clearly shows that both spectra
differs from each other. Based on this difference we may say that even
the best representation of our inhomogeneous source by the calculated
effective values in a homogeneous approach fails to reproduce the
observed spectrum.  The figure also shows the distribution of all
voxels of magnetic field inside the loop, viewing angle, nonthermal
electron number density and ambient density. One thing that strongly
support the use of 3D modeling is the viewing angle that dances over
the distribution plot under azimuth rotation or changes in
heliographic coordinates when maintaining all the rest equal. As the
gyrosynchrotron emission strongly depends on viewing angle, the
spectra changes accordingly.
 
\subsection{Flux dependencies on the central angle and source size}

\begin{figure}
 \begin{center}
    \FigureFile(80mm,100mm){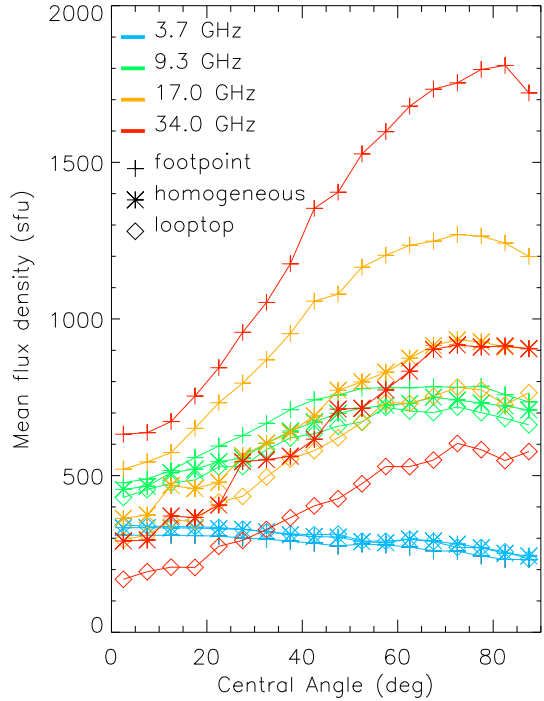}
 \end{center}
  \caption{Averaged flux densities of the model bank versus center angle.}
\label{fig:c2l}
\end{figure}

Although the model bank was constructed only over observed positions for 
the sake of usability of the models with observations and limitation of the 
processing time, we observed a general tendency of lower fluxes in the
disk center. To obtain a homogeneous distribution of models in center angle 
we distribute the models in steps of $5 ^ \circ $. Out of this distribution we observed 
84 models in the first channel (center angle from 0 to $5^ \circ $) that was calculated 
over seven observed positions for different flare parameters. Extracting 84 models from 
each other channel of the distribution randomly selected we averaged the calculated flux densities 
over the distribution for each frequency of observation. 
We repeated the process 100 times with different random sequences to average the fluxes 
in each channel. The result is shown in the figure \ref{fig:c2l}. The decrease in flux of the 
optically thin emission is clear. The averages of the flux density of our model bank are higher than the averages of 
observed fluxes, but the shown trend result in negligible fluxes for faint flares yielding to 
a decrement in the registered flares in any case. We also observe that footpoint concentrations show higher 
flux dependences in center angle, as expected from the gyrosynchrotron dependence on viewing angle. 

\begin{figure}
 \begin{center}
    \FigureFile(80mm,80mm){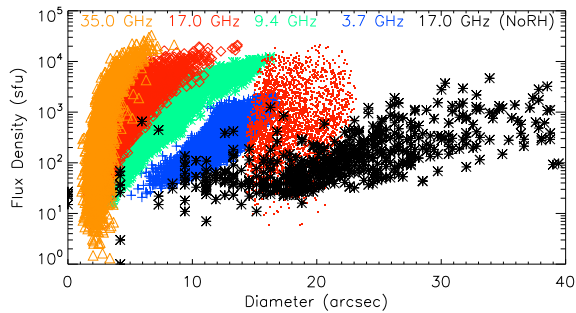}
 \end{center}
 \caption{Integrated flux density versus effective diameter of the
   emitting area. The red points are the effective diameter for 17 GHz
   (red diamonds) after convolution with a Gaussian beam of 15 arcsec}
\label{fig:area}
\end{figure}

In the figure \ref{fig:area} we see the effective diameter for
different frequencies and flux distribution. We can see that it
covers the variation of 17 GHz fluxes but not the size. The flux
variation for the four frequencies shown matches the observed range by
NoRP. In red diamonds we see 17 GHz from the images with 0.5 arcsec
resolution and the points after a convolution by a gaussian beam with
15 arcec. It is clear that our data bank with fixed loop size fails to
represent the observed sizes of 17 GHz (asterisks). We still have to
enlarge our loops to account for the bigger sources observed. Also, the 
dependence of observed fluxes at 17 GHz on size (\ref{fig:area}) is 
a characteristic that was not clearly reproduced by our model bank.

\subsection{Comparing results}
\label{cr}

\begin{figure*}
 \begin{center}
    \FigureFile(160mm,110mm){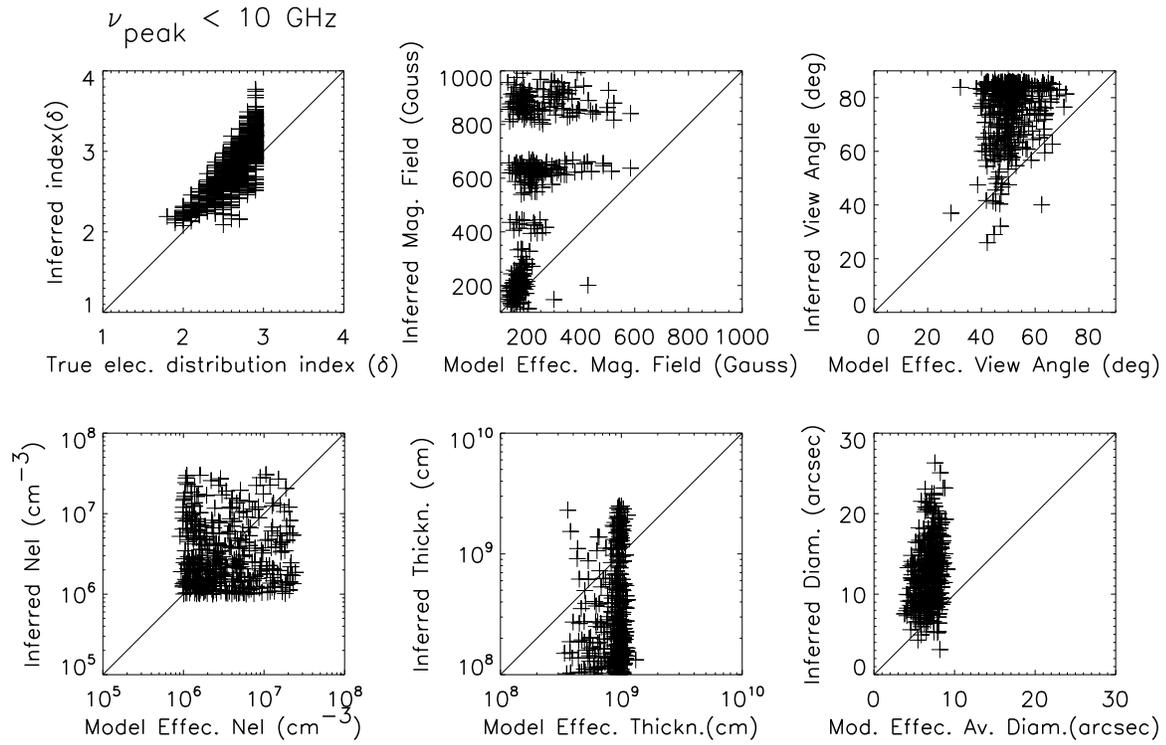}
 \end{center}
  \caption{Comparison between homogeneous and inhomgeneous modeling.}
\label{fig:differences}
\end{figure*}

A comparison of the homogeneous source approach with the actual source
characteristics of the emission showed important insights.  The figure
\ref{fig:differences} shows a comparison between the effective
parameters that describe our data cubes and parameters inferred by a
least square fit of a homogeneous source spectrum in the calculated
spectrum.  We selected from our model bank 909 spectra with peak
frequency ($\nu_{peak}$) lower than 10 GHz. We limited the sample to
maintain the 17/34 GHz emission in the optically thin regime to infer
the spectral index $\delta$ from Dulk's law between radiation spectral
index and the electron spectral index \citep{1985ARA&A..23..169D}.

\begin{figure*}
 \begin{center}
    \FigureFile(160mm,80mm){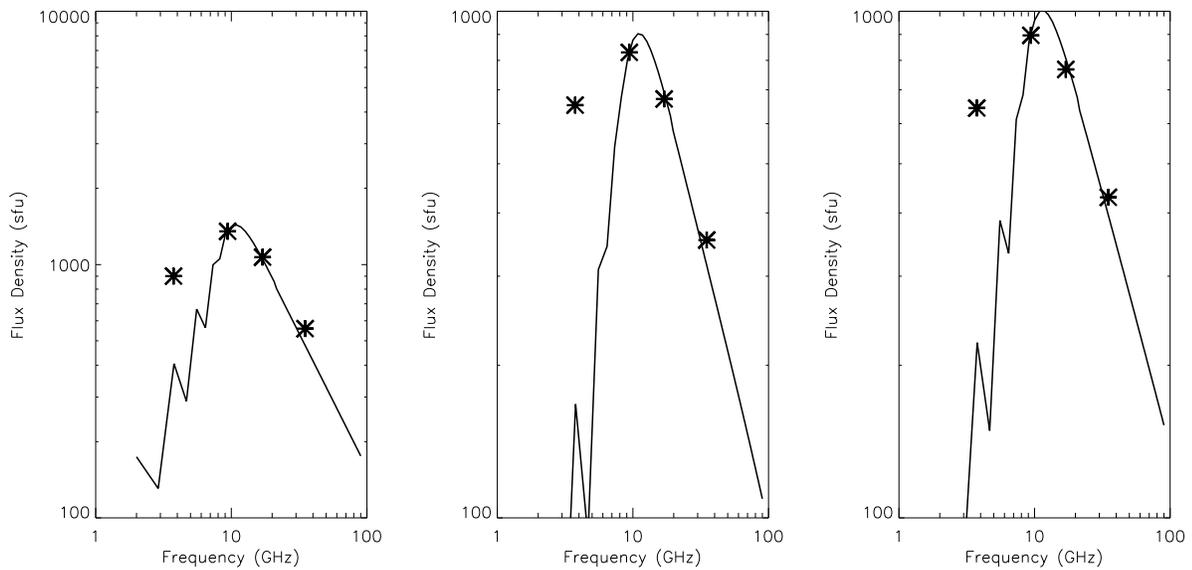}
 \end{center}
  \caption{Spectra fitting of homogeneous source in inhomogeneous data source.}
\label{fig:spectra}
\end{figure*}

We used a genetic algorithm \citep{1995ApJS..101..309C} to fit
homogeneous spectra on the inhomogeneous spectra data calculated in
our data bank. We implemented the genetic algorithm as a robust method to search 
for the region in the global parameter space associated with the maximum of a figure of 
merit related to the goodness of the fitting.  The goodness of fitting is measured by 
the maximum of  $1/\chi^2$. $\chi^2$ is the summation of squared differences between the 
observed and calculated flux densities. Some examples of the fitting are shown in 
figure \ref{fig:spectra} for three different flares at peak time. 

As a homogeneous source model we used a
gyrosynchrotron code with four fixed parameters: homogeneous pitch
angle distribution, range of energy from 10 keV to 100 MeV, ambient
density equal $10^{9} cm^{-3}$  
and ambient temperature of $10^{8} K$, 
and six variable parameters: electron spectral index $\delta$,
nonthermal electron density $N_{el}$, viewing angle $\theta$, magnetic
field B, area and depth. The population of spectra calculated by the
genetic algorithm for each fit had 2100 members and the best fit was
then extracted from the minimum $\chi^2$.

The result can be seen in the figure \ref{fig:differences}. The X-axes
contain the effective parameters that represents the data cubes and
Y-axes the inferred parameters from homogeneous approach. The spectral
index $\delta$ is roughly similar the index used but the rest differs
by orders of magnitude. Some of them are overestimated and some are
underestimated.

\subsection{Some observations}
To show the roughly estimation of the characteristics of our data bank
we got two flares observed by Nobeyama (NoRP and NoRH) and selected
from the data bank the best model to represent both flares by the
spectrum and brightness distribution. The number of models calculated
is still very modest but the result is quite promising.

\begin{figure*}
 \begin{center}
   \FigureFile(70mm,55mm){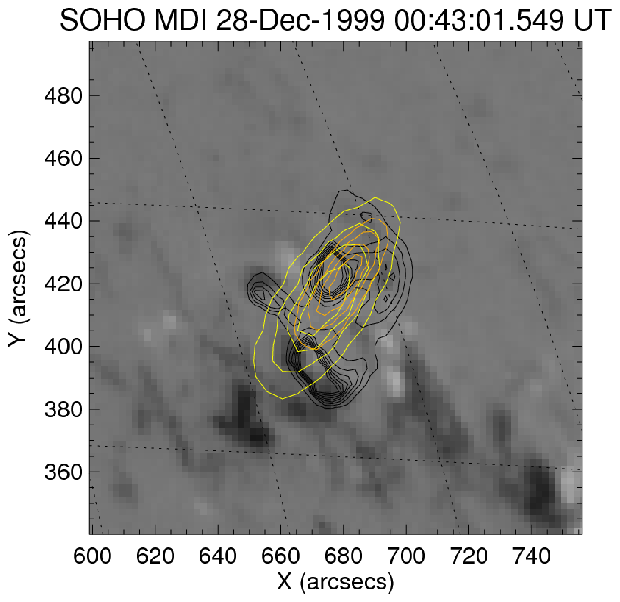}
   \FigureFile(70mm,55mm){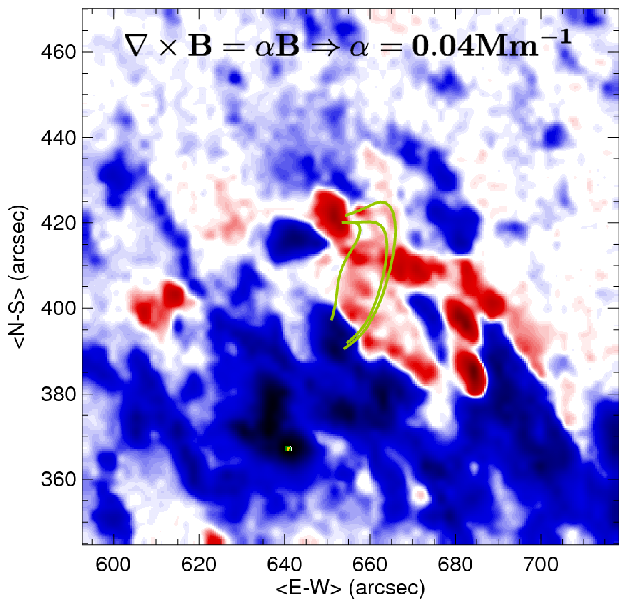}
   \FigureFile(70mm,55mm){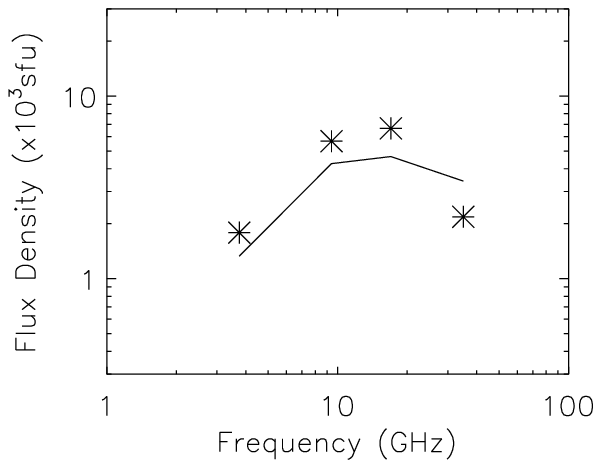}
   \FigureFile(80mm,62.86mm){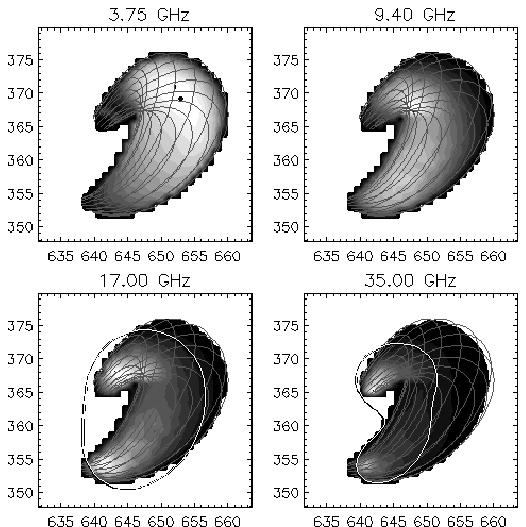}
 \end{center}
 \caption{Nobeyama flare $19991228\_0043$. Part {\bf a} shows the SDO
   magnetogram with the contours of 17 GHz in green and 34 GHz in
   yellow over-plotted. In {\bf b} is the magnetogram with some
   extrapolated linear force free field lines. In {\bf c} is the
   observed spectrum (asterisks) and the model spectrum (line). In
   {\bf d} is the model brightness distribution with field lines
   over-plotted and contours in 17 and 34 GHz after gaussian beam
   convolution (15 and 8 arcsec, respectively). The SOHO is a project 
of international cooperation between ESA and NASA.}
\label{fig:19991228}
\end{figure*}

Figure \ref{fig:19991228} shows the December 28, 1999 at 00:43 UT
flare with the contour plot of the 17/34 GHz observations over-plotted
on the magnetogram observed by MDI on board of SOHO satellite in the
part {\bf a} of the figure. Looking inside our data bank we find the
best fit model to the spectrum and image of the observed flare as
is shown in parts {\bf c} and {\bf d}. In part {\bf c} of the figure
some magnetic field lines are plotted. The field lines were
extrapolated solving the force free field equation with linear
solutions. The $\alpha=0.04$ Mm$^{-1}$ current parameter was selected to match
the orientation of the center line of the loop model. The extrapolated
line is $1.7 \times 10^{9} cm$ long and is roughly equal the length of
the loop central line with $2 \times 10^{9} cm$. The magnetic field at
the foot point of this extrapolated line is 1156 Gauss and the maximum
magnetic field of the model loop is 1366 Gauss.

\begin{figure*}
 \begin{center}
   \FigureFile(70mm,55mm){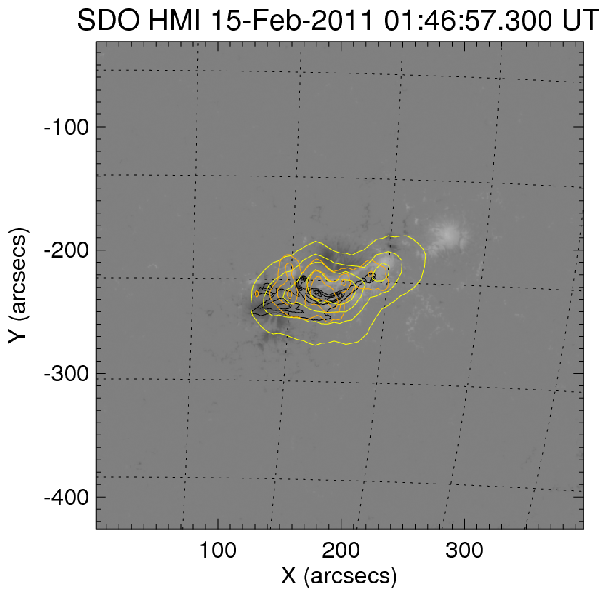}
   \FigureFile(70mm,55mm){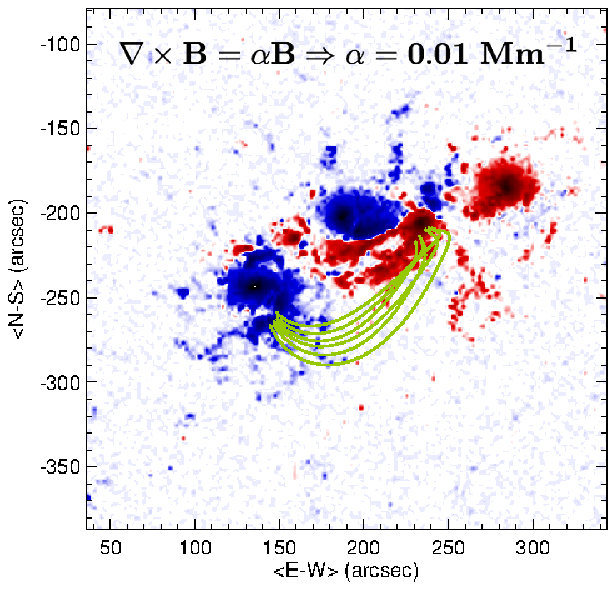}
   \FigureFile(70mm,55mm){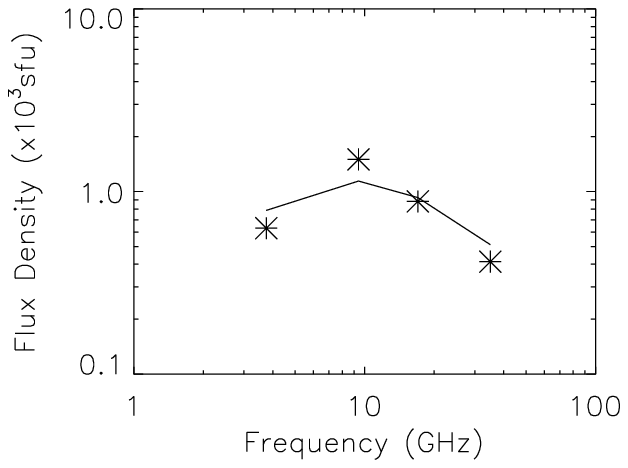}
   \FigureFile(80mm,62.86mm){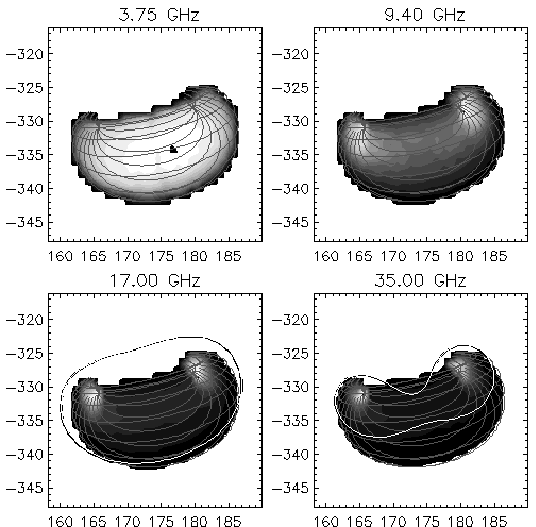}
 \end{center}
 \caption{Nobeyama flare $20110215\_0154$. Same as figure
   \ref{fig:19991228}. The SDO images are a 
courtesy of NASA/SDO and the AIA, EVE, and HMI science teams.}
\label{fig:20110215}
\end{figure*}

Figure \ref{fig:20110215} shows another observed burst in February
15, 2011 at 01:46 UT.  The shown magnetorgram is an observation of HMI on board of SDO satellite. 
The extrapolated field lines in this case is $6.5 \times 10^{9} cm $ long and the model loop length  is $2 \times 10^{9}
cm$. The footppoint magnetic field of the line is 1341 Gauss and the
footpoint magnetic field at the loop model is 1518 Gauss.


\section{Discussion and conclusions}

The missing flares of the Nobeyama sample at the Sun center are likely to 
flux densities below the instrument sensitivity of the less intense flares that are missed. 
This is clearly due to the flux density dependence on the viewing angle 
(as shown in our figure \ref{fig:c2l}). Thus, the geometry 
of the source may play an important role in the emission as was reviewed here. One 
important aspect of this is the comparison of the analysis of a flare that was produced 
inside an inhomogeneous environment and analyzed latter as a simplistic homogeneous 
source (as shown in the subsection \ref{cr}). Our intent was 
not to show different methods of homogeneous flare analyzes based 
in the spectrum fit but to use one method and evaluate the risk of getting misleading 
results. The genetic algorithm was chosen to be independent of initial guesses and for being 
robust to search for a global maximum for the figure of merit in the parameters space. It is 
clear that the homogeneous source hypothesis is not
appropriate to the possible flare geometry.  

The data bank of the models increases our capacity of flare analysis with a 3D loop model
and understanding of flare properties. We mention, for example, that in the figure \ref{fig:area} 
the roughly 
correlation of the observed flux density with source diameter at 17GHz is not  
obtained by the models with fixed loop size. Even if the model bank includes larger 
loops this may result in a broad range of flux versus diameter instead of a correlation. This 
property will be analyzed in a future tuning of the model bank. The first analysis of NoRP
and NORH flares is encouraging as shown by figures \ref {fig:19991228} and 
\ref{fig:20110215}. These analyzes were done by searching 
inside spectra bank given the flare heliographic position. This is an IDL routine that 
searches inside a structured variable with all parameters of the models. With the best 
fits of the spectra one may select the model with best the azimuth for the Nobeyama Radioheliograph 
observed map. 

Final refinements can be achieved in the future work calculating new models using different 
azimuth and loop size or any other flare parameters such as the plasma density and magnetic field strength. 
Taking into account the Razin effect may be important for the flare analysis. It is known that Razin 
supression plays significant role at lower microwave frequencies, $\nu<10$~GHz, for a large percentage of flares ($>50$\%)
observed with the spectrometer of Owens Valley Solar Array \citep{Melnikov et al:2008}. Strong Razin supression  was also detected 
with NoRH even at frequency 17~GHz but only for some flares \citep{Melnikov et al:2005, KuznetsovMelnikov:2012}. 
It means  that the higher ratios $n_0/B$ may happen in real solar flaring loops than it is taken in our model.

For the two flares shown in figures  \ref {fig:19991228} and \ref{fig:20110215} no refinement was made, thus 
they reflect properties of the actual model bank. They were chosen because one of us 
(T.S.N.Pinto) is analyzing the possibility to use force free extrapolation to render the loop 
volume for another publication.

\bigskip
\section{Acknowledgements}
We would like to thank the Nobeyama Radioheliograph, which is operated by the NAOJ/Nobeyama Solar Radio Observatory. 
For two of us (PJAS and VFM), this research was partly supported by the Maria Curie Exchange Program grant MC FP7-PEOPLE-2011-IRSES-295272. Also, PJAS gratefully acknowledges financial support by the European Commission through 
HESPE (FP7-SPACE-2010-263086). TSNP visited NSRO with support from the Japanese International Cooperation Agency (JICA)

\end{document}